\documentclass[fleqn]{2020SCGE}
\begin{document}

\ensubject{subject}

\ArticleType{Article}
\Year{2020}
\Month{January}
\Vol{63}
\No{1}
\DOI{??}
\ArtNo{000000}
\ReceiveDate{???}
\AcceptDate{???}

\title{Search for gamma-ray line signals around the black hole at the galactic center with DAMPE observation}{Search for gamma-ray line signals around the black hole at the Galactic center with DAMPE observation}

\author[1]{Tian-Ci Liu}{}%
\author[1]{Ji-Gui Cheng}{}
\author[1]{Yun-Feng Liang}{liangyf@gxu.edu.cn}
\author[1]{En-Wei Liang}{lew@gxu.edu.cn}%

\AuthorMark{Liu T C}

\AuthorCitation{Liu T C, Cheng J G, Liang Y F, et al}

\address[1]{Laboratory for Relativistic Astrophysics, Department of Physics, Guangxi University, Nanning 530004, China}


\abstract{The adiabatic growth of a black hole (BH) may enhance the dark matter (DM) density surrounding it, causing a spike in the DM density profile.
The spike around the supermassive BH at the center of the Milky Way may lead to a dramatic enhancement of the gamma-ray flux of DM annihilation from the galactic center (GC).
{In this work, we analyze the gamma-ray data of the innermost region (i.e., the inner 1$^\circ$) of the GC to search for potential line-like signals from the BH spike.
Such line-like signals could be generated in the process of DM particles annihilating into double photons. We adopt the gamma-ray data from the Dark Matter Particle Explorer (DAMPE).
Although the DAMPE has a much smaller effective area than the Fermi-LAT, the gamma-ray line search can benefit from its unprecedented high energy resolution. }
No significant line-like signals are found in our analysis. We derive upper limits on the cross section of the annihilation based on this non-detection.
We find that despite the DAMPE's small effective area for photon detection, we can still place strong constraints on the cross section ($\left<\sigma v\right>\lesssim10^{-27}\,{\rm cm^3\,s^{-1}}$) in the spike scenario due to the very bright model-expected flux from the spike. Our results indicate that either DM does not annihilate primarily through the $\gamma\gamma$ channel in the mass range we considered or no sharp density spike is present at the GC.}

\keywords{Dark matter, $\gamma$-ray telescopes, Black holes, Galactic center}

\PACS{95.35.+d, 95.55.Ka, 97.60.Lf, 98.35.Jk} 

\maketitle


\begin{multicols}{2}
\section{Introduction}
Many astrophysical and cosmological phenomena indicate the presence of a large amount of dark matter (DM) particles in the Universe \cite{dmrv1,dmrv2}. If the DM consists of weakly interacting massive particles (WIMPs), they may annihilate and generate observable gamma-ray/cosmic-ray signals.
One interesting case is that DM particles directly annihilate into monochromatic gamma-rays, i.e., via $\chi\chi \rightarrow {\gamma}\gamma$, which generates an emission line over the background at gamma-ray energies.
This line signal is thought to be the smoking gun of
\Authorfootnote

\noindent
a new physics since normal astrophysical processes are not expected to generate this signal.
With the gamma-ray data from EGRET \cite{pullen07egret}, Fermi-LAT \cite{atwood09lat}, and the Dark Matter Particle Explorer (DAMPE) \cite{dampe}, line-like signals have been widely studied for various searching targets \cite{fermi10line1,fermi12line2,bringmann130gev,weniger130gev,gs12dsphline,huang12130gev, tempel130gev,bloom13earthlime,fermi13line3,fermi14line4,fermi15line5, anderson16gcline,liang16line,ls19line,szq21line}.
Tentative line signals have also been reported from the galactic center (GC) \cite{bringmann130gev,weniger130gev} or galaxy clusters \cite{liang16line,szq21line}, although a robust detection is still absent.
The non-detection often leads to strict constraints on the annihilation cross section or the lifetime of DM annihilation / decay into double photons.

The adiabatic growth of a black hole (BH) increases the DM density near the BH and forms a spike in the density profile \cite{Gondolo99spike,Ullio01spike,Merritt03spike,Gnedin03spike}.
The resulting density spike leads to a strong enhancement of the DM annihilation signal \cite{fields14,johnson19pwave,jcg2020ucmh,ygw21m87,xzq21m31}.
Since the presence of a supermassive black hole (SMBH) Sgr A* at the center of the Milky Way is well-confirmed \cite{Schodel02gc16yr,Ghez08gcorbit}, its influence must be considered when searching for DM signals from the GC. If the SMBH Sgr A* at the GC grew adiabatically from an initial seed in the NFW DM halo, then the DM profile near the BH steepened to a spike with a sharp halo index of $\gamma_{\rm sp}>2$ \cite{fields14}. The spike radius is expected to be on the order of $\sim1$ pc, corresponding to an angular size of $\sim0.01^\circ$, well below the resolution of a gamma-ray telescope. The DM annihilation from the spike will appear as a point source in the gamma-ray sky.

In this work, we use DAMPE gamma-ray data to search for line-like signals from the innermost ($<1^\circ$) region of the GC.
Previously, no study dedicated to line-like signals has concentrated on this small region.
We consider the scenario of a DM density spike caused by the adiabatic growth of the SMBH Sgr A* at the GC.
We derive how the spike J-factor relies on DM particle parameters (DM mass $m_\chi$ and cross section $\left<\sigma v\right>$), and point out that when $\left<\sigma v\right>$ is small, the presence of a BH spike will lead to greater enhancement of the J-factor.
Therefore, considering the spike is especially important for a line-like signal, which is expected to have a small cross section theoretically \cite{Bergstrom:1997fh,Ullio:1997ke,Chalons11nmssm,tempel130gev,Chen:2013bi,feng2016} and experimentally \cite{fermi15line5}.
We use three years of publicly available DAMPE data and adopt analysis methods optimized for line signals.
Because of the very high energy resolution, the DAMPE is well suited for searching for monochromatic line-like gamma-ray signals.
We obtain strong constraints on the annihilation cross section of the DM gamma-ray line in the spike scenario.

\section{Gamma rays from the DM spike at the GC}
\label{sec:flux}
The expected gamma-ray flux from the DM annihilation can be expressed as \cite{charles16review}
\begin{equation}
\frac{d\phi}{dE} = \frac{1}{4\pi}\frac{\langle{\sigma}v\rangle}{2m_{\chi}^2}\frac{dN_{\gamma}}{dE}{\times}J_{\rm ann},
\label{equ_dmflux}
\end{equation}
where $m_\chi$ is the DM mass, $\langle{\sigma}v\rangle$ is the velocity-averaged annihilation cross section, and $dN_\gamma/dE$ is the differential gamma-ray yield per annihilation.
The second term in Eq.(\ref{equ_dmflux}) is the so-called J-factor, which depends on the DM density distribution $\rho(r)$ within the source
\begin{equation}
J_{\rm ann} = \int\int_{\rm los}{\rho^2(r)}dsd\Omega.
\end{equation}
The expected gamma-ray flux is proportional to the J-factor.

The DM distribution in the presence of a spike follows the density profile \cite{fields14}
\begin{equation}
\rho(r) =
\begin{cases}
0, & \quad r<4 M \\
\frac{\rho_{\mathrm{sp}}(r) \rho_{\mathrm{ann}}}{\rho_{\mathrm{sp}}(r)+\rho_{\mathrm{ann}}},& 4 M \leq r<R_{\rm sp} \\
\rho_{b}(R_{\rm sp} / r)^{\gamma_{c}},& \quad R_{\rm sp} \leq r<R_{H} \\
\rho_{H}(R_{H} / r)^{\gamma_{H}},& \quad R_{H} \leq r
\end{cases}
\label{eq:profile}
\end{equation}
The third and fourth segments represent the smooth DM halo of the Milky Way.
If $\gamma_{c}=1$ and $\gamma_{H}=3$, it is the commonly used Navarro-Frenk-White (NFW \cite{nfw96nfw}) density profile.
$\rho_{b}$ and $\rho_{H}$ are the densities at radius $R_{\rm sp}$ ($R_{\rm sp}=0.2r_h$, see below) and $R_{H}$, respectively, which can be determined according to the local DM density $\rho_\odot$ (i.e., $\rho_{b}=\rho_\odot(r_\odot / R_{\rm sp})^{\gamma_{c}}$, $\rho_{H}=\rho_\odot(r_\odot / R_{H})^{\gamma_{c}}$).
The Sun is located at $r_\odot=8.5\,{\rm kpc}$ from the GC. In our calculation, we adopt $R_{H}=20\,{\rm kpc}$ \cite{fermi13line3} and the local DM density $\rho_\odot=0.4\,{\rm GeV/cm^3}$ \cite{fermi13line3,Catena10localdm,Salucci10localdm}.

The presence of the SMBH changes the DM density inside the radius of the gravitational influence of the SMBH, $r_{h}=GM / v_{0}^{2}$, with $M$ the mass of the BH and $v_{0}$ the velocity dispersion of DM.
The impact on the DM density is visible in the first and second segments of Eq. (\ref{eq:profile}).
The DM density will be dramatically enhanced by the adiabatic growth of the BH, and a very steep density profile forms (i.e., the spike).
For a spike of collisionless DM, we have $\rho_{\rm sp}(r)\propto r^{-\gamma_{\rm sp}}$, with $\gamma_{\rm sp}=\left(9-2 \gamma_{c}\right) /\left(4-\gamma_{c}\right)$ \cite{Gondolo99spike,Ullio01spike,Merritt03spike,Gnedin03spike}.
For the canonical NFW DM halo of the Milky Way, it results in $\gamma_{\rm sp}=7/3$.
In the inner part of the spike, a saturation density is reached because of the annihilation of the DM particles, $\rho_{\rm ann} \equiv m_{\chi} /(\langle\sigma v\rangle {t})$. The saturation density relies on the mass of the DM particle $m_\chi$, the annihilation cross section $\left<\sigma v\right>$, and the time over which annihilations have occurred (approximately the age of the BH, $t = t_{\rm BH}$).
For $r<2R_s$, we simply set $\rho(r)=0$, where $R_s=2M$ is the Schwarzschild radius.
The related parameter values adopted in this work are \cite{fields14}: $M=4\times10^6\,{\rm M_\odot}$, $v_0=105\pm20\,{\rm km/s}$, and $t_{\rm ann}=10^{10}\,{\rm yrs}$.

\section{Searching for line-like signals with DAMPE data}
\label{sec:data}
\subsection{DAMPE data reduction}
The DAMPE is a space-based GeV-TeV cosmic-ray and gamma-ray detector that can measure gamma-rays with unprecedentedly high energy resolution ($<1.5\%$ for $\gtrsim10$ GeV, 68\% containment) \cite{dampe}. The scientific goals of the DAMPE include measuring cosmic-ray spectra with exceptional energy resolution and energy reach, studying high-energy gamma-ray astronomy, and detecting indirect DM signals.
The DAMPE is sensitive to gamma rays in the energy range from $>$1~GeV to $\sim$10~TeV \cite{xu18photon,duan19dmpst}.
Because of its very high energy resolution, the DAMPE is well suited for searching for monochromatic line-like gamma-ray signals.
The DAMPE team has used DAMPE data to search for line-like gamma-ray signals from four optimized regions of interest (ROIs) \cite{dampe21line,xzl22linesel}.
It is shown that although the DAMPE has an effective area much smaller than Fermi-LAT, its observation can give even stronger constraints on the DM parameters for the search with a large ROI.

\begin{figure}[H]
\centering
\includegraphics[width=0.45\textwidth]{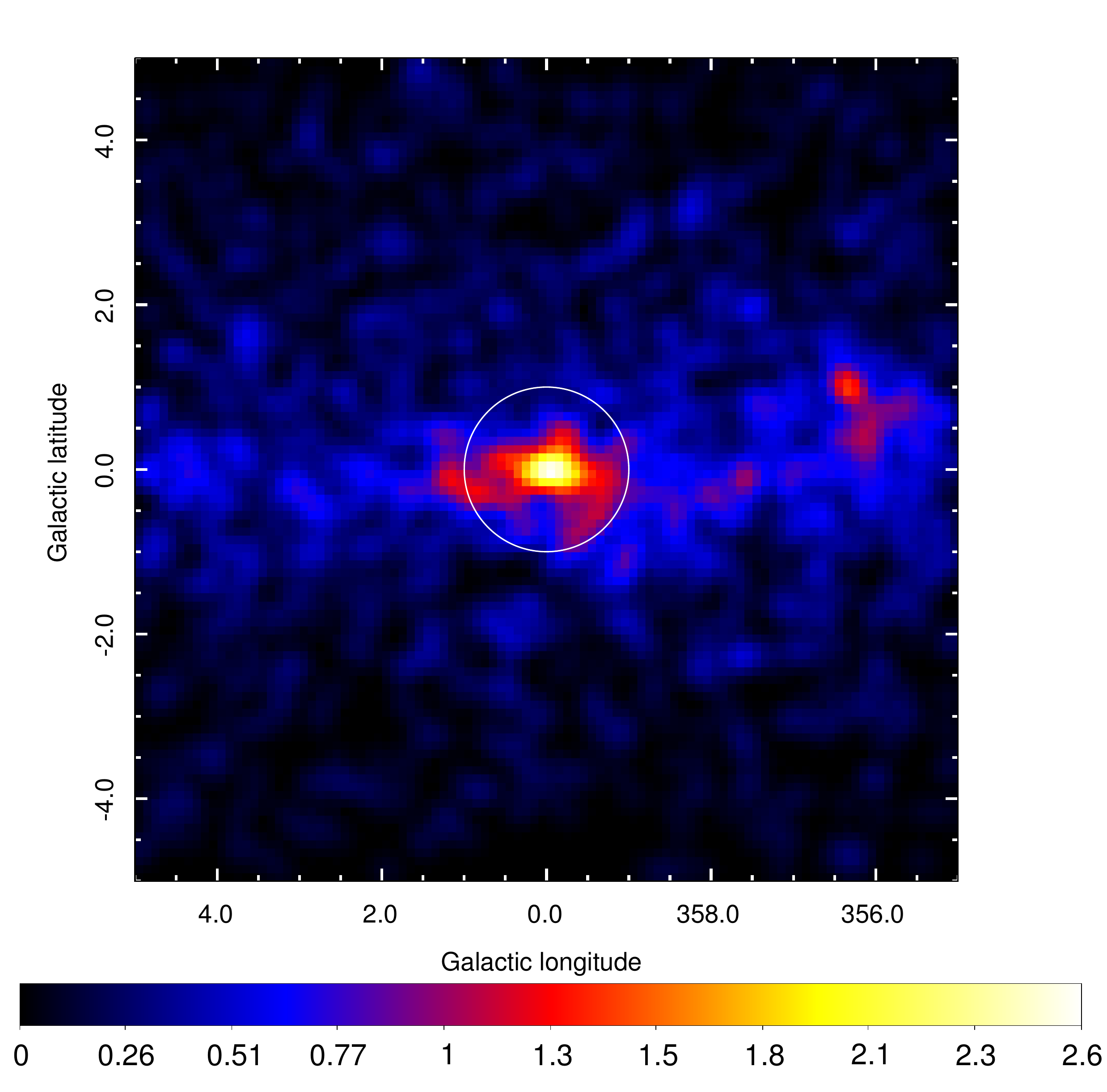}
\caption{DAMPE 3-3000 GeV counts map toward the GC. The white circle encloses the region where we search for gamma-ray line signals.}
\label{fig:cmap}
\end{figure}

\begin{figure}[H]
\centering
\includegraphics[width=0.45\textwidth]{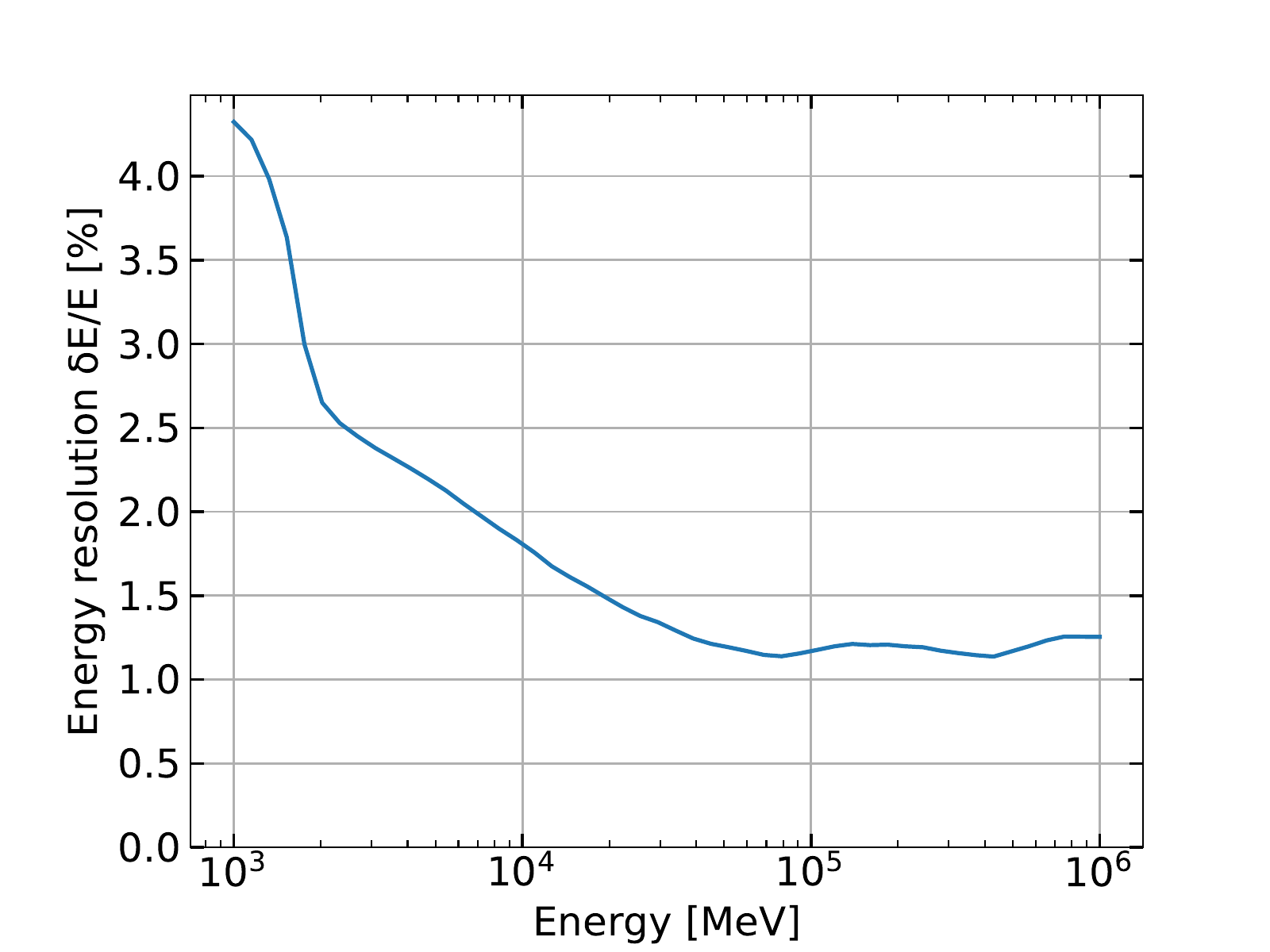}
\caption{Exposure-weighted energy resolution (68\% containment half-width) of the DAMPE for the observation data toward the GC.
}
\label{fig:eres}
\end{figure}

In this work, we use the publicly available data of the DAMPE from 2016 January 1 to 2018
December 31 (i.e., MET 94608000 - MET 189302399)\footnote{\url{https://dampe.nssdc.ac.cn/dampe/dataquerysc.php}}. We select the photons 1$^\circ$ around the Sgr A* (${\rm RA}=266.417^\circ$, ${\rm Dec}=-29.008^\circ$)\footnote{\url{http://simbad.u-strasbg.fr/simbad/}} between 3 and 1000 GeV.
We adopt the default {\tt evtype=2} data, which include both LET and HET data \cite{duan19dmpst}.
To ensure data quality, only photons with $0.5\leq\cos(\theta)\leq1$ are adopted, where $\theta$ is the incidence angle in the detector frame. The DAMPE observation of the GC for 3-1000 GeV photons is shown in Figure \ref{fig:cmap}.
We use {\tt DmpST} software (version: 1.2.0)\footnote{\url{https://dampe.nssdc.ac.cn/dampe/dampetools.php}} to prepare the materials used in the following line search analysis.
We use the {\tt Events} submodule to select data within a 1$^\circ$ radius of Sgr A* and calculate the exposures toward this source using the {\tt Exposure} submodule.
In addition, an exposure-weighted energy resolution curve for the direction of the GC (Figure \ref{fig:eres}) is generated using the {\tt PlotWedisp\_E} function in the {\tt Exposure} module.
The subsequent analysis is based on these ingredients (the selected photon list, the exposures, and the energy resolution) with our own code.

\subsection{Line-like signal search}
To search for spectral lines, we use an unbinned likelihood analysis in sliding energy windows \cite{bringmann130gev, weniger130gev,fermi13line3,fermi15line5,liang16line}.
For a series of line energies $E_{\gamma}$ from 6~GeV to 660~GeV\footnote{We use DAMPE data of 3-1000 GeV. However, considering the window size, we only search for DM in the mass range from 6 GeV to 660 GeV.}
with increments of 0.5 ${\sigma}_E(E_{\gamma})$, the unbinned fitting is performed in a small energy window of $(E_{\gamma}-0.5E_{\gamma},~E_{\gamma}+0.5E_{\gamma})$.
The $E_{\gamma}$ is the energy of the line signal that is fixed in the fit of each window, and ${\sigma}_E(E_{\gamma})$ is the energy resolution ($68{\%}$ containment half-width of the reconstructed incoming photon energy) of the DAMPE at $E_{\gamma}$.
In each window, we approximate the background spectrum (null model) as a single power law, which is reasonable considering the small size of the energy windows.
The signal model (alternative model) is described by the function of a power-law background plus a line component.
Considering the energy dispersion, the line component is, in fact, the DAMPE energy dispersion function \cite{duan19dmpst} rather than a delta function. In this work, we use a Gaussian function to approximate the energy dispersion of the DAMPE with the Gaussian $\sigma$ being the 68\% energy resolution in Figure \ref{fig:eres}.

The likelihood functions for the null model and the signal model are
\begin{equation}
\ln \mathcal{L}_{\text {null }}= \sum_{{i}=1}^{N} \ln \left(F_{\mathrm{b}}\left(E_{{i}}\right) {\epsilon}\left(E_{{i}}\right)\right) \\
-\int F_{\mathrm{b}}(E){\epsilon}(E) {\rm d} E,
\end{equation}
and
\begin{equation}
\begin{aligned}
\ln \mathcal{L}_{\text {sig }}
=& \sum_{{i}=1}^{N} \ln \left[F_{\mathrm{b}}\left(E_{{i}}\right) {\epsilon}\left(E_{{i}}\right)+F_{\mathrm{s}}\left(E_{{i}}\right) {\epsilon}\left(E_{\gamma}\right)\right] \\
&-\int\left[F_{\mathrm{b}}(E) {\epsilon}(E)+F_{\mathrm{s}}(E) {\epsilon}\left(E_{\gamma}\right)\right] dE,
\end{aligned}
\end{equation}
where $F_{\rm b}$ and $F_{\rm s}$ are the power-law background and the line component, respectively, and $\epsilon(E)$ is the DAMPE exposure toward the GC.
Maximizing the likelihoods of the two models gives the best-fit parameters and the test statistic (TS) of the line component, ${\rm TS}\triangleq 2(\ln{\mathcal{L}_{\rm sig}}-\ln{\mathcal{L}_{\rm null}})$.
The local significance is the square root of the TS according to the asymptotic theorem of Chernoff \cite{Chernoff1954}. The global significance is obtained after considering the trial factor, which would decrease the significance.
The analysis method for the search of line-like signals with the gamma-ray data has been widely described in the literature. We refer readers to Refs. \cite{weniger130gev,fermi13line3,fermi15line5,liang16line} for details.
For instance, Sec. II C and Sec. III A of \cite{liang16line} present the unbinned likelihood method and sliding window technique, respectively.

\section{Results}
\label{sec:rs}
\begin{figure}[H]
\centering
\includegraphics[width=0.45\textwidth]{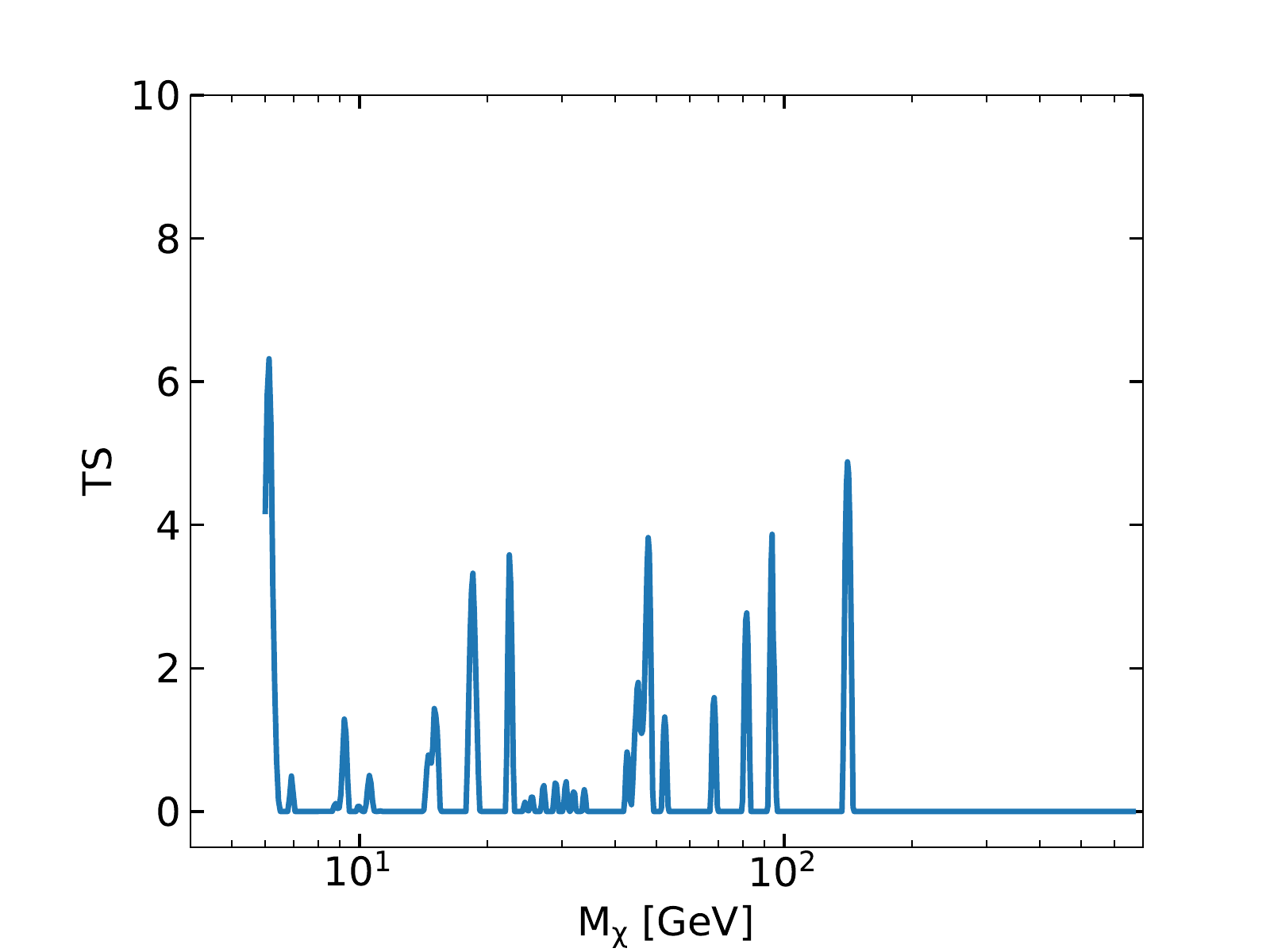}
\caption{TS value as a function of DM mass (i.e., line energy) in the sliding window analysis. No significant (TS$>25$) signals are found for all the DM masses that we have examined.}
\label{fig:ts}
\end{figure}

The results of the search for line-like signals from the most inner (1$^\circ$) region of the GC are shown in Figure \ref{fig:ts}, where we demonstrate how the TS value of the putative line-like signal varies as a function of the DM mass. No significant (i.e., TS$>25$, corresponding to a local significance of $5{\sigma}$) line signals are found for all the energies from 6 GeV to 660 GeV.
The comb-like structures in the TS-$m_\chi$ curve are mainly due to statistical fluctuations or systematic effects.
The most significant line-like excess appears at $\sim6.2\,{\rm GeV}$ with a TS value of $\sim$ 6.3.
This low TS value is far from the criterion of a detection (${\rm TS}\sim25$).

Since no significant line-like signals are found, we place constraints on the flux of the putative gamma-ray line and the cross section for DM annihilation into two photons $\langle \sigma v \rangle_{\chi\chi\rightarrow \gamma\gamma}$. To derive a 95\% confidence level upper limit on the line flux at a given $m_\chi$, we vary the prefactor of the line and find the value at which the log-likelihood is smaller by 1.35 compared to the best-fit model.
{Because of the impact of the point-spread function (PSF) of the instrument, the photons from the central source may spill over the 1$^\circ$ ROI, especially for relatively low energies. This PSF effect needs to be corrected. The flux UL given by the likelihood analysis is divided by a factor $C(E,1^\circ)$ to obtain the realistic UL of the line signal, where $C(E,1^\circ)$ is the PSF containment within the 1$^\circ$ circle. We can use {\tt DmpST} to calculate $C(E,1^\circ)$, which is energy dependent, ranging from 89\% at 6 GeV to 96\% at 600 GeV.}
The resulting flux upper limits of the putative line at 95\% CL are at a level of $(1-10)\times10^{-10}\,{\rm ph\,cm^{-2}\,s^{-1}}$.
The flux UL is then converted to constraints on the DM parameters.

\begin{figure*}[t]
\centering
\includegraphics[width=0.45\textwidth]{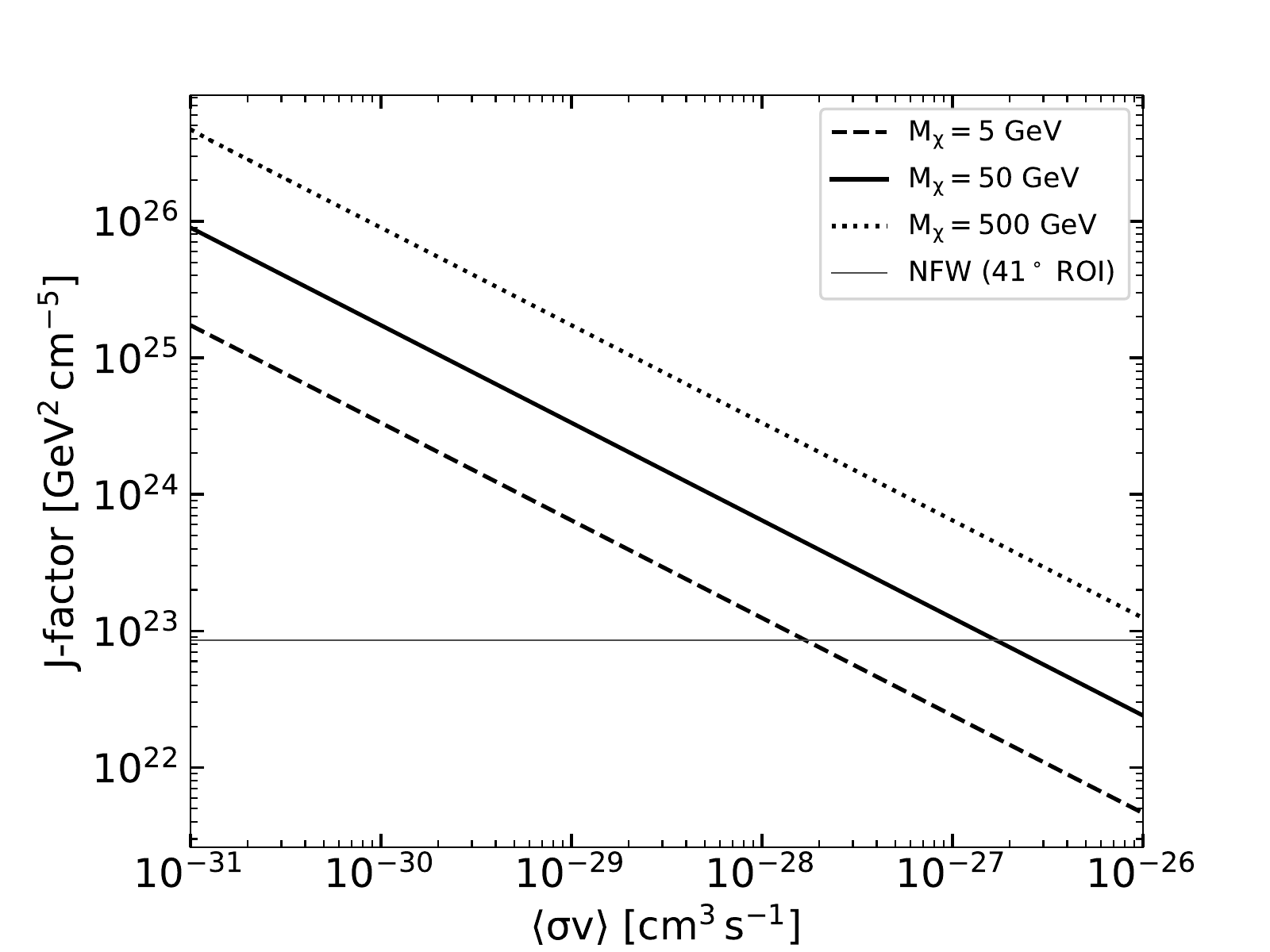}
\includegraphics[width=0.45\textwidth]{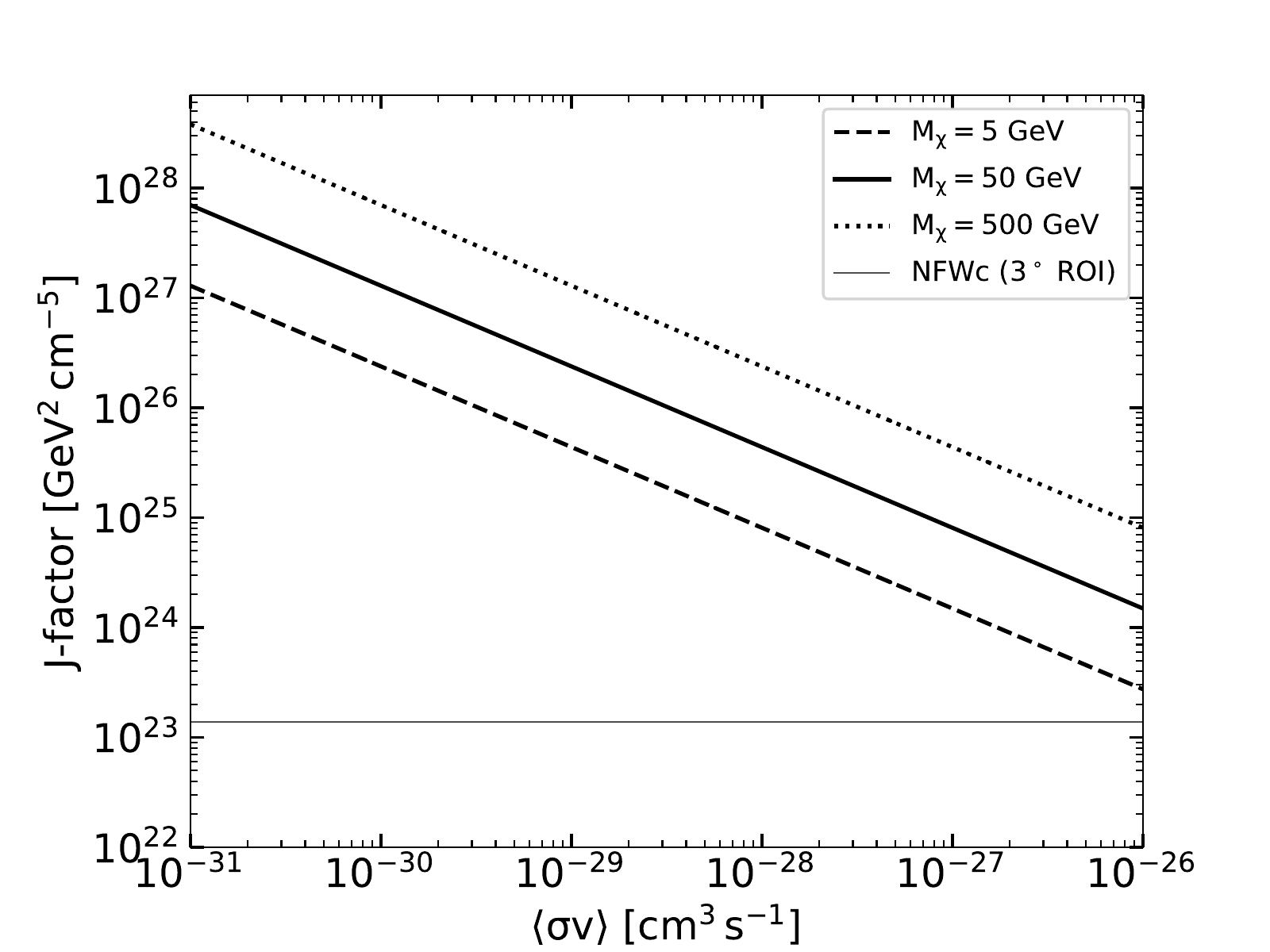}
\caption{The J-factor of a BH spike depends on $\left<\sigma v\right>$ and $m_\chi$ through the saturation density $\rho_{\rm ann}$. Here, we show how the J-factor varies as a function of $\left<\sigma v\right>$ for 5, 50, and 500 GeV DM. The left panel is for an NFW distribution of the smooth DM halo of the Milky Way and the right is for an NFWc DM distribution.}
\label{fig:jf}
\end{figure*}

\begin{figure*}[t]
\centering
\includegraphics[width=0.6\textwidth]{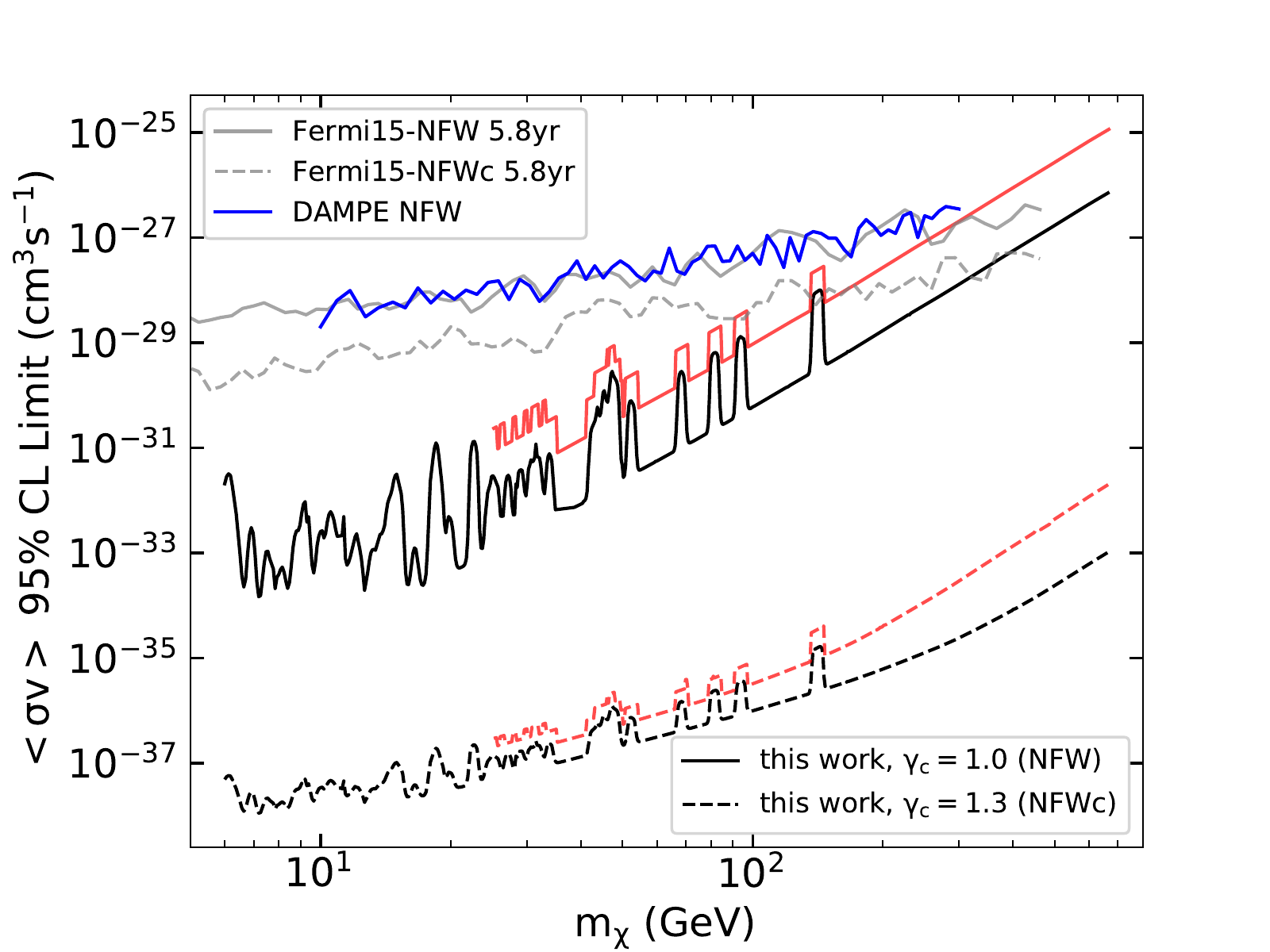}
\caption{The 95\% confidence level upper limits on the DM cross section $\langle \sigma v \rangle_{\chi\chi\rightarrow \gamma\gamma}$ in the BH spike scenario.
The black lines are results based on a sliding-window unbinned analysis, which is commonly used in the searches for gamma-ray lines.
At large $m_\chi$, the upper-limit curve is smooth without fluctuations because no photons are observed within the corresponding energy window. The two curves correspond to different halo indices $\gamma_c$. The red lines are the conservative upper limits based on an inclusive spectrum of the GC 1$^\circ$ region. {Also shown are the results from Fermi-LAT \cite{fermi15line5} and DAMPE \cite{dampe21line} collaborations in the {\it no-spike} scenario. We see that the presence of a BH spike will highly enhance the model-predicted flux of the DM annihilation, and we, therefore, can obtain very strong constraints from the non-detection.}
}
\label{fig:ul}
\end{figure*}

For the $\chi\chi\to\gamma\gamma$ annihilation channel, the expected gamma-ray flux is given by
\begin{equation}
S_{\rm line}(E) = \frac{1}{4\pi} \frac{\left<\sigma v\right>_{\chi\chi\to\gamma\gamma}}{2m_\chi^2} \ 2\delta(E-E_{\rm line}){\times}J_{\rm ann},
\label{eq:lfl}
\end{equation}
where $E_{\rm line}=m_\chi$ is the energy of the emitted monoenergetic photons. The gamma-ray flux limits can be converted to the constraints on the DM annihilation cross section $\left<\sigma v\right>_{\chi\chi\to\gamma\gamma}$ using Eq. (\ref{eq:lfl}) after the J-factor $J_{\rm ann}$ is determined.

For normal DM profiles such as NFW \cite{nfw96nfw} and Einasto \cite{einasto}, if the profile parameters are given, the J-factor of the DM source is established and does not change with the annihilation cross section $\left<\sigma v\right>$ and DM mass $m_\chi$.
However, the J-factor of the BH spike depends on $\left<\sigma v\right>$ and $m_\chi$ through the saturation density $\rho_{\rm ann}=m_\chi/(\left<\sigma v\right>t)$.
A larger $\left<\sigma v\right>$ implies that the spike has a smaller $\rho_{\rm ann}$ and that the saturation occurs at a larger radius, which will significantly reduce the annihilation rate in the central part of the halo.
Conversely, a smaller $\left<\sigma v\right>$ will result in a very high $\rho_{\rm ann}$. Because of the steep profile of the spike, the central saturation part contributes significantly to the annihilation gamma-ray flux.
In Figure \ref{fig:jf}, we show how the J-factor varies with the annihilation cross section for 5, 50, and 500 GeV dark matter. Only when the cross section is larger than the thermal relic value \cite{Steigman12thermalcs}, the total DM flux (proportional to the J-factor) from the NFW halo is expected to higher than that radiated from the BH spike.
It is worth noting that, the existing results without considering the spike have constrained the cross section $\left<\sigma v\right>_{\chi\chi\to\gamma\gamma}$ to a level of $10^{-26} - 10^{-30}\,{\rm cm^{3}\,s^{-1}}$ \cite{fermi15line5}, meaning that the enhancement with the spike considered will be very large.
As a comparison, for the continuum emission from DM (e.g., the $b\bar{b}$ or $\tau^+\tau^-$ channel), the current best constraints from gamma rays are approximately $\gtrsim10^{-26}\,{\rm cm^3\,s^{-1}}$; thus, adding the spike contribution would not substantially improve the results. Namely, considering the presence of the BH spike is more important for the analysis of spectral lines.

In Figure 5 we present the resulting constraints on $\langle \sigma v \rangle_{\chi\chi\rightarrow \gamma\gamma}$ for the DM masses from 6~GeV to 660~GeV.
The spike power-law $\gamma_{\rm sp}$ relies on the initial density profile of the halo where the BH is embedded.
Here, we consider two DM distributions of the smooth halo of the Milky Way, a canonical NFW ($\gamma_c$=1.0) and an adiabatically contracted NFW (NFWc, $\gamma_c$=1.3).
Figure 5 shows that, because the presence of a BH spike significantly enhances the expected J-factor, we can give quite a strong restriction on the annihilation cross section in this scenario.
Our analysis places upper limits on the cross section down to the levels of $\sim10^{-33}-10^{-27}\,{\rm cm^3\,s^{-1}}$ (for NFWc, it can be limited to $\sim10^{-37}-10^{-33}\,{\rm cm^3\,s^{-1}}$), largely stronger than those without the consideration of the spike \cite{fermi15line5,dampe21line}.
Such strong constraints would pose a significant challenge to relevant models.

{For the above results, we use Chernoff's theorem \cite{Chernoff1954} to determine the upper limits (i.e., corresponding to the flux at which $\Delta\ln\mathcal{L}$ is varied by -1.35 compared to the maximum). However, at high energies, the availability of the asymptotic theorem is not guaranteed since the gamma-ray counts are limited. In \cite{dampe21line}, the DAMPE collaboration avoids this issue by requiring at least 30 photons in each energy window. 
Since we are using a very small ROI, adopting the same strategy restricts the highest line energy in our analysis to only 25 GeV. For this reason, we alternatively use an inclusive spectrum to derive more robust/conservative upper limits for $m_\chi>25$ GeV.
In Figure~\ref{fig:ul}, the constraints based on the inclusive spectrum are also shown (dashed lines). To derive these constraints, we require the model-expected photon counts ($N_{\rm exp}$) of the line signal satisfying $F(x=N_{\rm obs};u=N_{\rm exp})<0.05$, where $F(x;u)$ is the cumulative distribution functions of the Poisson distribution, and $N_{\rm obs}$ is the number of photons from the direction of 1$^\circ$ around Sgr A* in the energy range of ($E_{\rm line}-5{\sigma}_E$, $E_{\rm line}+5{\sigma}_E$).
With this conservative method, we obtain upper limits weaker by a factor of a few, which, however, do not influence the main conclusions of our work.
}

{We need to clarify that there exists, in fact, an upper boundary of the exclusion region (see also \cite{ygw21m87}). Since the saturation density is proportional to $\left<\sigma v\right>^{-1}$, for a very large annihilation cross section, the spike density in the most central region becomes very small. The expected gamma-ray fluxes from the annihilation within the spike should also be small. Therefore, a very large cross section cannot be excluded by our analysis. We find that for all the DM masses we considered and for both $\gamma_c=1.0$ and $1.3$, the upper boundary of the exclusion region is $>10^{-18}\,{\rm cm^3\,s^{-1}}$, much larger than the current upper limits based on the observations of the Galactic halo \cite{fermi15line5} and dSphs \cite{Liang:2016bxu}. Thus, for better visualization, we do not plot the upper boundary in the figures, except for the results in Section \ref{sec:br}.}

\section{Discussion}

\subsection{Implications for DM models}

{If the presence of the spike is confirmed in the future by, for example, stellar kinematics, then the results of our work can be used to identify DM models.

A tentative line was found at $\sim$43 GeV in the Fermi-LAT observation of 16 nearby galaxy clusters, with a global significance of $3.0\sigma$ \cite{liang16line}. Although the line is found to have a strange temporal behavior of its significance \cite{szq21line}, the interesting possibility of a signal from DM annihilation has not been excluded yet. This gamma-ray line signature can be interpreted by the model with a boson propagator and triangle charged particle loop \cite{Bergstrom:1997fh}, the semiannihilation dark matter model \cite{DEramo:2012fou}, and so on \cite{feng2016}. The required cross section of $\left<\sigma v\right>\sim5\times10^{-28}\,{\rm cm^3\,s^{-1}}$ is however not compatible with the BH spike. 

In the GC, there is a GeV excess that may be from the dark matter annihilation \cite{Hooper:2013rwa,Zhou:2014lva,Fermi-LAT:2017opo}. Recently, Abdughani et al. \cite{Abdughani:2021pdc} proposed a common origin of the GeV excess, the anti-proton excess, and the g-2 anomaly in the framework of the NMSSM and claimed a dark matter particle mass of $\sim60\,{\rm GeV}$. In this case, the velocity-averaged cross section of annihilation into the gamma-ray line is from $10^{-33}$ to $10^{-30}\,{\rm cm^3s^{-1}}$ \cite{Chalons11nmssm}. Therefore, at least for $\gamma_c=1$, the constraints shown in Figure \ref{fig:ul} are not tight enough to exclude the presence of a dark matter spike. }

\subsection{Uncertainties related to the spike parameters}
\begin{figure*}[t]
\centering
\includegraphics[width=0.45\textwidth]{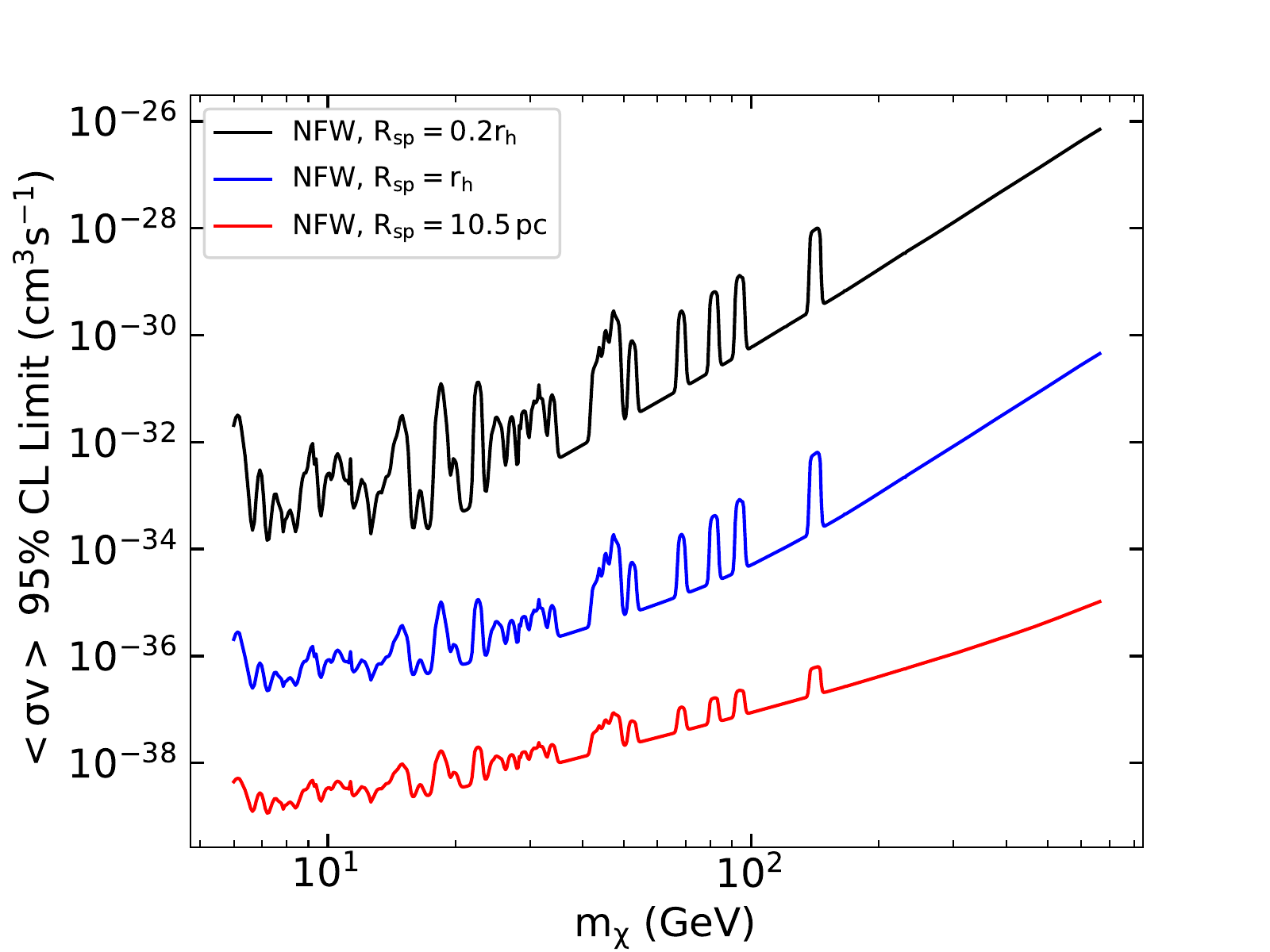}
\includegraphics[width=0.45\textwidth]{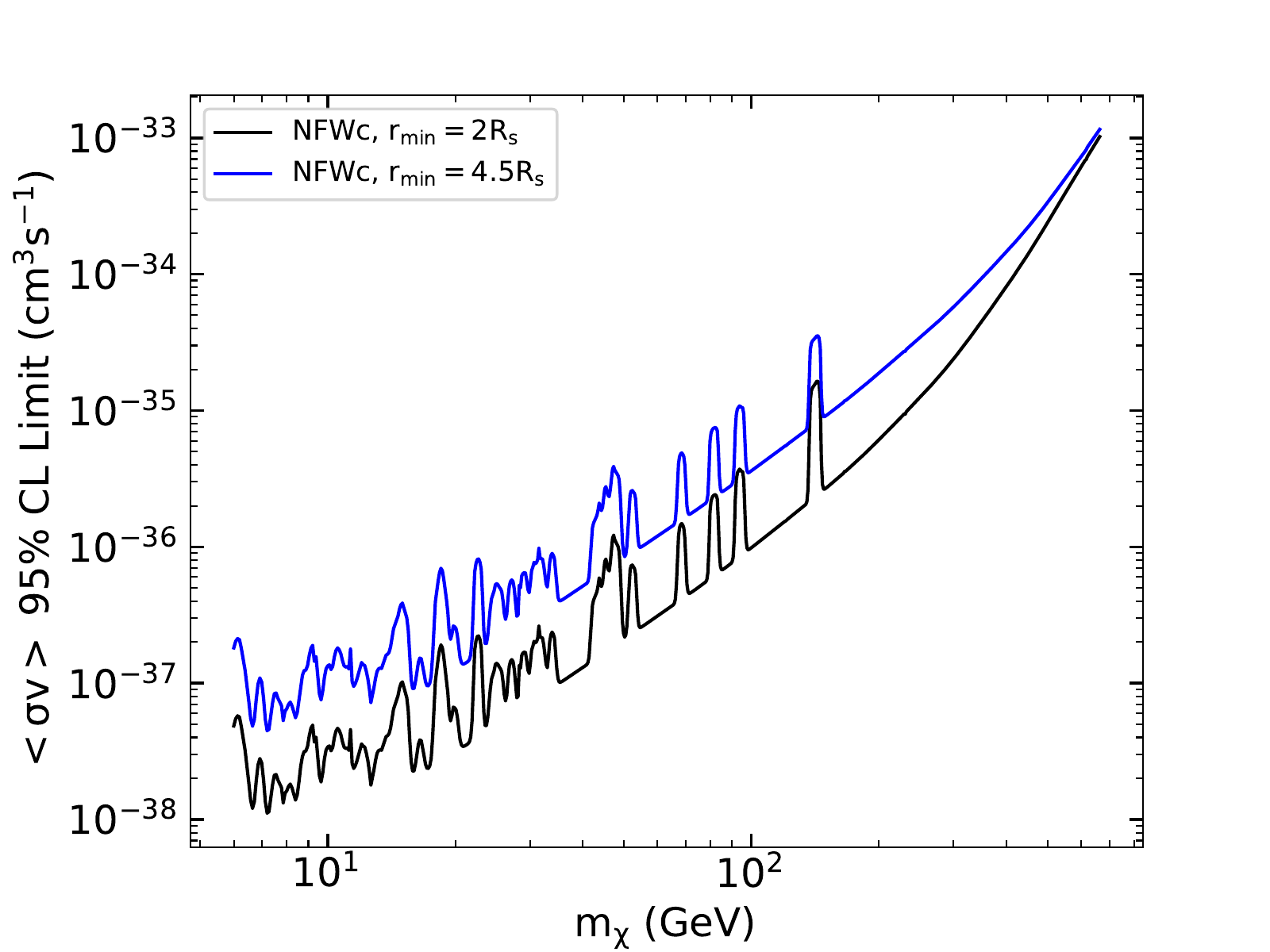}
\includegraphics[width=0.45\textwidth]{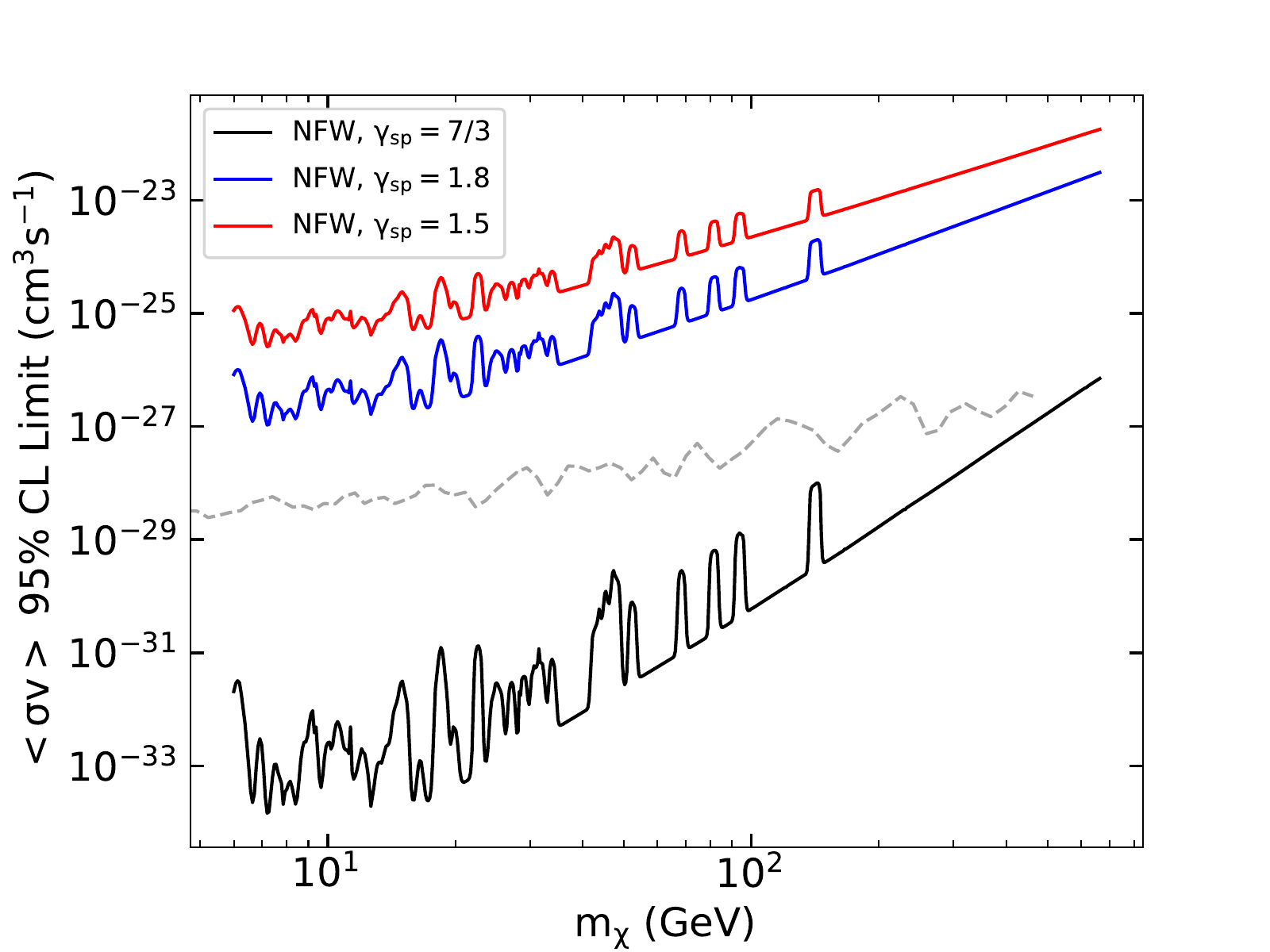}
\includegraphics[width=0.45\textwidth]{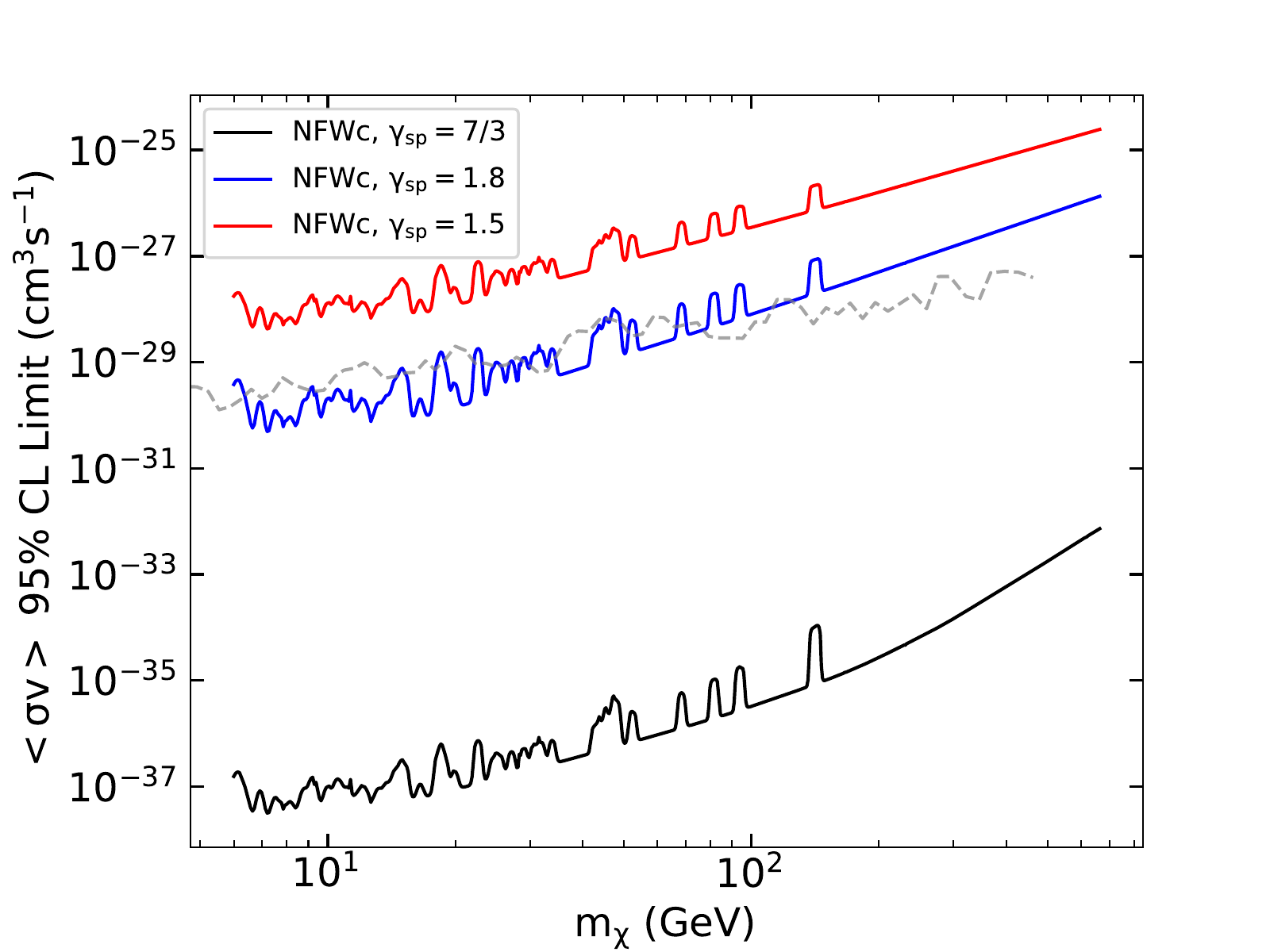}
\caption{These plots demonstrate the upper limits on the DM cross section to gamma-ray lines for different spike parameters: spike radius $R_{\rm sp}$ (top-left panel), inner truncation radius $r_{\rm min}$ (top-right panel), and spike slope $\gamma_{\rm sp}$ (bottom panels). The black lines in all the plots are the benchmark results that have been shown in Figure \ref{fig:ul}. The gray lines in the two bottom panels are the updated constraints using the Fermi-LAT data \cite{fermi15line5} in the {\it no-spike} scenario.}
\label{fig:3fig}
\end{figure*}

{For this investigation, the profile of the dark matter spike is crucial.
Here, we further discuss how the parameters of the profile will affect the constraints.

{\it Spike radius $R_{\rm sp}$.} We notice that different choices of the $R_{\rm sp}$ are reported in the literature \cite{Gondolo99spike,Merritt03spike,fields14,Lacroix18}, and the value of $R_{\rm sp}=0.2r_h\simeq0.3\,{\rm pc}$ adopted in this work is conservative. This choice is also consistent with the recent constraints by the GRAVITY collaboration according to the orbit of the star S2 \cite{Gravity20}, which restricted $R_{\rm sp}$ to smaller than {10\,{\rm pc}} for an initial NFW dark matter distribution. Here, we also present results based on other values of the spike radius, $R_{\rm sp}=1.6\,{\rm pc}$ (i.e., $r_h$), and $R_{\rm sp}=10.5\,{\rm pc}$. The latter one is calculated with $R_{\mathrm{sp}}\sim\alpha_{\gamma} r_{H}({M}/{\rho_{H} r_{H}^{3}})^{1/{(3-\gamma_c)}}$ \cite{Gondolo99spike}. The results are shown in the top left panel of Figure \ref{fig:3fig}. We find that by taking larger $R_{\rm sp}$ values, the constraints could be stronger by 3-6 orders of magnitude.}

{\it Inner truncation radius $r_{\rm min}$.} {In the above calculation, we set the inner truncation radius of the spike to $r_{\rm min}=2R_s$, which is the marginally bound orbit of the BH. However, for the innermost region (near the horizon of the black hole), the dark matter distribution is not well understood. We alternatively try to use the innermost stable circular orbit ($3R_s$ for a Schwarzschild BH and $\leq4.5R_s$ for a rotating BH) as the $r_{\rm min}$. Since the spin information of the SMBH at the GC is unclear, we conservatively adopt $r_{\rm min}=4.5R_s$. We find that for $\gamma_c=1.0$, this only affects the results at a level of $<10^{-3}$.
For the case of $\gamma_c=1.3$, using $r_{\rm min}=4.5R_s$ will weaken the constraints on the cross section by a factor of a few (top-right panel of Figure \ref{fig:3fig}).
The reason is that the smaller $\left<\sigma v\right>$ for the NFWc spike corresponds to a higher saturation density $\rho_{\rm ann}$, and therefore the contribution from the most central region becomes more nonnegligible.}

{\it Spike slope $\gamma_{\rm sp}$}. {The most important parameter that may affect the model-predicted flux from the spike is the $\gamma_{\rm sp}$. The benchmark results (Figure \ref{fig:ul}) rely on the spike with $\gamma_{\rm sp}>7/3$.
Here, we show how the results will be if no strong spike is presented.
We adopt the slopes of $\gamma_{\rm sp}=1.5$ and $\gamma_{\rm sp}=1.8$ to derive the constraints (bottom panels of Figure \ref{fig:3fig}).
We can see that such soft profiles lead to a substantial weakening of the limits.}

\subsection{Possible caveats of the presence of a steep spike}
{We have shown that the stringent constraints and related conclusions we obtained above strongly depend on the presence of a steep spike. However, the formation of such a steep profile is speculative. In Ref. \cite{Ullio01spike}, it is argued that a steep spike requires unusual initial conditions for the Galactic halo and the SMBH.
The BH must grow adiabatically from a tiny initial mass and be precisely at the center of the DM halo.
In a realistic scenario, there are many possibilities to weaken the spike \cite{fields14,Ullio01spike}.
(i) Forming a spike requires the DM particles to be initially cold near the BH,
but baryonic effects (such as gravitational scattering off stars) will heat the DM particles and lead to a decrease in $\gamma_{\rm sp}$.
Depending on the time scale of the DM heating, the resulting profile has $\gamma_{\rm sp}\sim1.8$ for 10 Gyr of DM heating and $\gamma_{\rm sp}=1.5$ if $>20$ Gyr.
(ii) The SMBH must grow adiabatically from a small mass, and the growth time scale should be long compared to $10^7$ years. In the opposite case, if the BH appears suddenly (possibly associated with subhalo or BH mergers), a shallow spike with $\gamma_{\rm sp}=4/3$ is expected.
If the seed BH is instantaneously formed and then grows adiabatically, the spike will have an intermediate slope.
(iii) The BH growth had to occur within the inner 50 pc of the halo center for the spike to arise.
The offset of the seed BH from the center of the DM distribution ($>50\,{\rm pc}$) may lead to a complete flattening of the spike unless a significant growth of the BH occurs after it reaches the halo center.}

\subsection{Branching ratio to the $\gamma\gamma$ channel}
\label{sec:br}
\begin{figure}[H]
\centering
\includegraphics[width=0.45\textwidth]{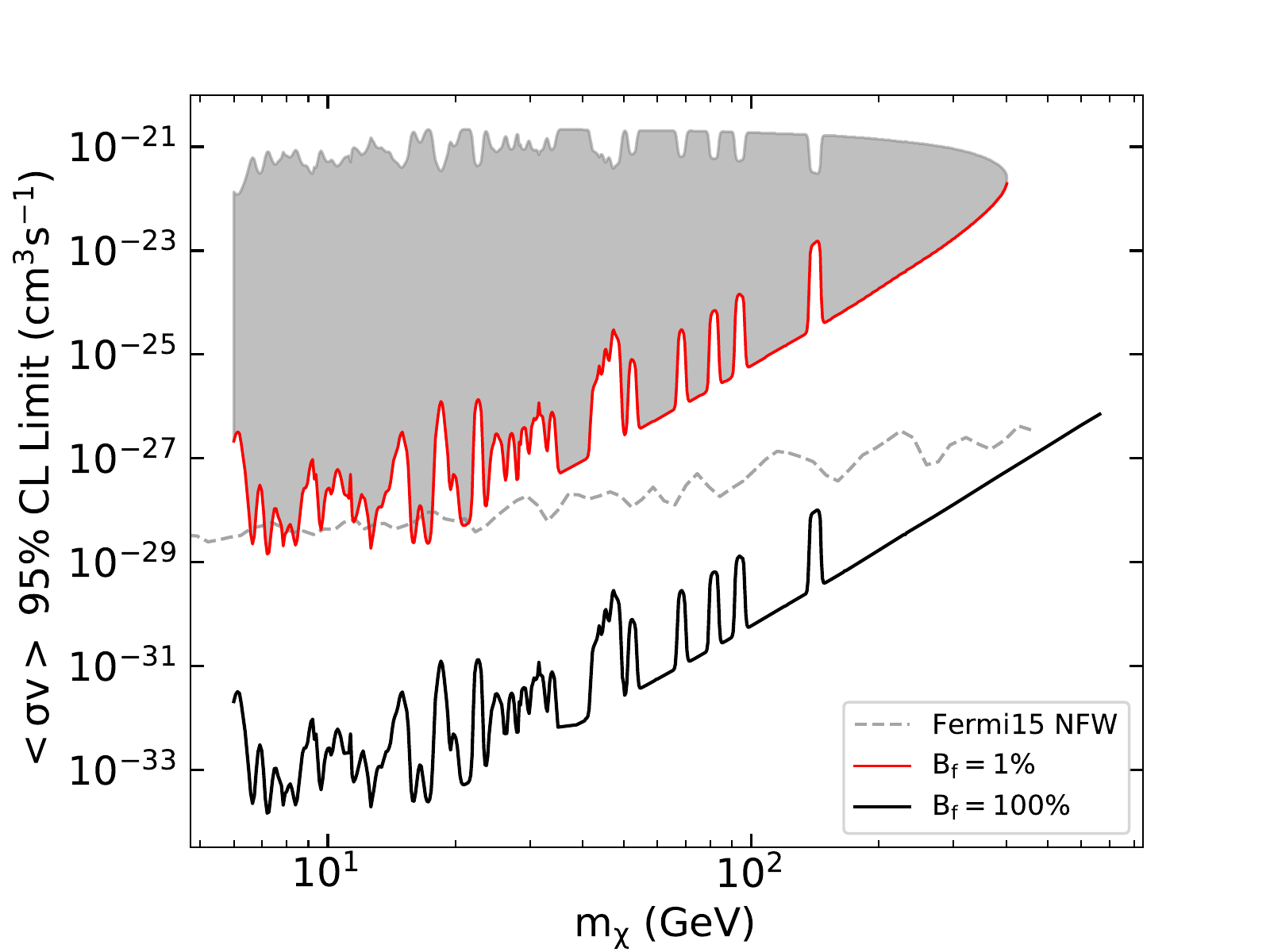}
\caption{Constraints on the cross section
assuming different branching ratios into photon final states. The black line is the benchmark result the same as that in Figure \ref{fig:ul}. Because of the existence of an upper boundary of the excluded parameters, only parameters in the gray region can be constrained for $B_\gamma=1\%$.}
\label{fig:br}
\end{figure}

{Above, we simply assume that DM particles annihilate entirely through the $\gamma\gamma$ channel. However, considering the realistic DM models, the branching ratio to $\gamma X$ final states is loop-suppressed, since electrically neutral DM particles are not directly coupled to photons. For WIMP DM, typical suppression for one-loop processes is on the order of $10^{-4}$ compared to the tree-level processes. The annihilation into monochromatic gamma rays usually has a very small branching ratio compared with the total annihilation cross section \cite{Bergstrom:1997fh,Ullio:1997ke,Chalons11nmssm,tempel130gev,Chen:2013bi,feng2016}.

In this case, because the large total cross section reduces the saturation density $\rho_{\rm ann}$, the expected line signal from the spike will be lowered so that the obtained constraints are largely weakened. For a branching ratio of $B_\gamma=10^{-4}$, we find that for all the DM masses we considered, the model-expected flux is always lower than the measured flux UL,
so the current analysis based on DAMPE observation cannot rule out any regions of parameter space.
For $B_\gamma=10^{-2}$, although some parameters can be excluded (Figure \ref{fig:br}), the constraints are substantially weaker than the benchmark results that assume a 100\% branching ratio into two photons.

However, it should be noted that in the case of a small $B_\gamma$, the DM continuum emission produced through other annihilation channels will exceed the fluxes of the observed gamma-ray point sources near the GC \cite{fields14}, so the total cross section and correspondingly the $\left<\sigma v\right>_{\gamma\gamma}$ can also be severely constrained.
Furthermore, there exist mechanisms (e.g., Sommerfeld enhancement \cite{Chen:2013bi} and internal bremsstrahlung \cite{bringmann130gev}) that can enhance the cross section to photons and make the annihilation directly to gamma-rays dominate the total cross section.}

\section{Summary}

Robust detection of a monochromatic gamma-ray line would be a smoking gun of the presence of particle DM. Great efforts have been made to search for such signals in various targets. In this work, we analyze the DAMPE data toward the GC and especially focus on the innermost 1$^\circ$ region. Our search results do not identify any significant line emissions in the energy range of 6$-$660~GeV. Thus, we set upper limits on the DM annihilation cross section to monochromatic gamma rays. Compared with the constraints of no spike \cite{fermi15line5}, our current limits are much tighter in a wide energy range.

Given that we have limited the DM annihilation cross section of photon final states down to $\sim10^{-33}\,{\rm cm^3\,s^{-1}}$ under the spike scenario, this suggests that either DM does not significantly annihilate through the $\gamma\gamma$ channel (unless the annihilation is dominated by higher partial wave processes, e.g., a p-wave or d-wave \cite{johnson19pwave}) or there is no sharp density spike at the GC.
For the latter case, possibilities include \cite{fields14} (i) the nonadiabatic growth of the SMBH (such as BH mergers), (ii) the heating of DM by gravitational scattering off stars, and (iii) DM particle self-interactions.
These mechanisms would flatten the spike and decrease the annihilation signals.
To summarize, since the presence of an SMBH at the center of the Milky Way is well established \cite{Schodel02gc16yr,Ghez08gcorbit}, the fact that no signal from the spike is found must be considered in constructing relevant models.

The results of this paper can be improved with the Fermi-LAT observation, which has a much larger acceptance for photon detection \cite{atwood09lat}.
For the line search analysis of a large ROI, the sensitivity of the DAMPE can be close to or even better than that of Fermi-LAT \cite{dampe21line} because of its unprecedentedly high energy resolution.
However, because we are focusing on a very small ROI ($1^\circ$), the advantage of the high energy resolution cannot take effect due to the poor statistics.
However, as seen above, our analysis with the DAMPE has already been able to provide valuable conclusions.

\Acknowledgements{We acknowledge data resources from DArk Matter Particle Explorer (DAMPE) satellite mission supported by Strategic Priority Program on Space Science, and data service provided by National Space Science Data Center of China.
This work is supported by the National Natural Science Foundation of China (Grant No.12133003, U1731239 and 11851304), Guangxi Science Foundation (grant No. 2017AD22006 and 2019AC20334) and Bagui Young Scholars Program (LHJ).}

\InterestConflict{The authors declare that they have no conflict of interest.}


\bibliographystyle{unsrt}
\bibliography{ref1}

%
%
%
%
%
%
%
%
%
%
%
%
%
%

\end{multicols}
\end{document}